\documentclass[a4paper]{jpconf}
\usepackage{graphicx}
\begin{document}
\title{Persistent current noise}

\author{Andrew G. Semenov$^{1}$
and Andrei D. Zaikin$^{2,1}$}

\address{$^{1}$ I.E.Tamm Department of Theoretical Physics, P.N.Lebedev
Physics Institute, 119991 Moscow, Russia\\
  $^{2}$ Institut f\"ur Nanotechnologie, Karlsruher Institut f\"ur
Technologie (KIT), 76021 Karlsruhe, Germany}

\ead{semenov@lpi.ru}

\begin{abstract}
We demonstrate that persistent current in meso- and nanorings may
 fluctuate down to zero temperature provided the current
operator does not commute with the total Hamiltonian of the
system. For a model of a quantum particle on a ring we explicitly
evaluate PC noise power which has a form of sharp peaks which become broadened
for multi-channel rings or in the presence of dissipation.
PC noise can be tuned by an external magnetic flux which is a fundamental
manifestation of quantum coherence in the system.
\end{abstract}

Persistent currents (PC) in normal meso- and nanorings pierced by
external magnetic flux is one of fundamental consequences of quantum coherence of electrons. While the average value of PC was intensively investigated in the literature during last decades, not much is known about equilibrium fluctuations od PC. At non-zero $T$
it is quite natural to expect non-vanishing thermal fluctuations of PC. But at $T \to 0$ the system approaches its (non-degenerate)
ground state and, hence, at the first sight no PC fluctuations would be possible. Below we will demonstrate that this naive conclusion is in general not quite correct: No PC fluctuations are expected in the zero temperature limit only provided the current operator commutes with the total Hamiltonian of the system, otherwise fluctuations of persistent current occur even in the ground state exactly at $T=0$.

Let us consider a simple model of a quantum particle with mass $M$ on a
1d ring of radius $R$ pierced by magnetic flux $\Phi_x$. The particle position on the ring is parameterized by the angle $\theta$, which is the quantum mechanical variable of interest in our problem. The Hamiltonian for this system reads
\begin{equation}
   \hat H=\frac{(\hat \phi +\phi_x)^2}{2MR^2}+U(\theta )+\hat H_{int}(\theta,X),
\end{equation}
where $\hat \phi =-i\frac{\partial}{\partial\theta}$ is the
(dimensionless) flux operator, $U(\theta)$ defines the potential
profile for our particle, $\phi_x=\Phi/\Phi_0$ and $\Phi_0$ is the
flux quantum. The collective variable $X$ represents all other degrees of freedom interacting with the particle. Note that within our model the interaction term involves only  the coordinate (not momentum) operator. Such models were previously studied in a
number of papers \cite{GZ98,Buttiker,Paco,GHZ,HlD,SZ} in the context of
various dissipative environments and, in addition, could be
of interest for the problem of PC in superconducting
nanorings in the presence of quantum phase slips \cite{AGZ}.

Let us define the current operator in the Schr\"odinger
representation
\begin{equation}
    \hat I=\frac{e}{2\pi}\dot{\hat \theta}=\frac{ie}{2\pi}[\hat H,\hat\theta ]=\frac{e(\hat\phi +\phi_x)}{2\pi
    MR^2}.
\end{equation}
In the Heisenberg picture the current operator is given by
standard formula $\hat I(t)=e^{it\hat H}\hat I e^{-it\hat H}$. Let us
also introduce  the equilibrium current-current correlation function
$\langle\hat I(t)\hat I(0)+\hat I(0)\hat I(t)
\rangle$ and define PC noise power
\begin{equation}
S(t)=\langle\hat I(t)\hat I(0)+\hat I(0)\hat I(t)
\rangle-\langle\hat I\rangle^2=\int\frac{d\omega}{2\pi}S_\omega e^{-i\omega t}.
\label{corrK}
\end{equation}
Employing the full set of eigenstates $\hat
H|m\rangle=\varepsilon_m(\phi_x)|m\rangle$ after a straightforward
calculation we obtain
\begin{equation}
S(t)=\frac{1}{\mathcal Z}\sum\limits_{m,n} e^{it(\varepsilon_m-\varepsilon_n)}|\langle m|\hat I|n\rangle|^2(e^{-\beta\varepsilon_m}+e^{-\beta\varepsilon_n})-2\langle\hat I\rangle^2
\end{equation}
and
\begin{equation}
 S_\omega=4\pi\mathcal P\delta(\omega)+\frac{2\pi}{\mathcal Z}\sum\limits_{m\neq n}|\langle m|\hat I|n\rangle|^2(e^{-\beta\varepsilon_m}+e^{-\beta\varepsilon_n})
\delta(\omega+\varepsilon_m-\varepsilon_n),
\label{so}
\end{equation}
where ${\mathcal Z}$ is the grand partition function for our system, $\mathcal P$ is some constant.

We observe that PC noise power has the form of peaks at frequencies
equal to the distance between the energy levels with non-zero matrix
elements of the current operator plus an additional peak at zero
frequency. In the zero temperature limit $T\rightarrow 0$ the amplitude of this peak tends to zero and Eq. (\ref{so}) reduces to
\begin{equation}
   S_\omega=2\pi\sum\limits_{m>0}|\langle m|\hat I|0\rangle|^2\left(\delta(\omega+\varepsilon_0-\varepsilon_m)+\delta(\omega+\varepsilon_m-\varepsilon_0)\right)
\end{equation}
This result demonstrates  that PC fluctuations indeed persist down
to $T=0$ in which case peaks of PC noise power $S_\omega$ occur at frequencies corresponding to transitions between energy levels for which the matrix elements of the current operator differ from zero. In other words, PC correlator differs from zero
dwn to zero temperature provided the current operator does not commute with the total Hamiltonian of the system. A specific feature of PC noise is the dependence of $S_\omega$ on the external magnetic flux $\phi_x$. This dependence occurs due to the presence of quantum coherence in the system and disappears if this coherence gets
destroyed. Hence, such sensitivity of PC noise spectrum to the flux can
be used as a measure of quantum coherence in our system.

It is important to emphasize that, while the above results hold for a single particle on a ring, in other situations PC noise spectrum might look differently. For instance, broadening of such peaks inevitably occurs in ensembles of
rings or individual rings with many conducting channels. In this case the
total PC noise produced by the system is given by the sum of a large
number of very close peaks effectively resulting in a much smoother and
broader noise spectrum. Likewise, in the presence
of dissipation due to interaction of the particle with other (quantum)
degrees of freedom the energy levels acquire a finite width, the peaks get broadened and the noise power should differ from zero also in a wider
range of frequencies. Two examples will be considered below.

 Our first example is a particle on the ring in periodical
potential. Let us employ the imaginary time technique and define the generating
functional
\begin{equation}
\mathcal Z[\zeta ]=\int_0^{2\pi}
d\theta_0\sum\limits_{m=-\infty}^\infty e^{2\pi im\phi_x}\int\limits^{\theta_0+2\pi m}_{\theta_0}\mathcal
D\theta e^{-\int\limits_0^\beta d\tau\left(\frac{MR^2\dot
\theta^2}{2}+U(\theta)-i\zeta\dot \theta\right)}. \label{genf}
\end{equation}
This functional enables one to evaluate all imaginary-time current
cumulants by taking the derivatives with respect to the source field  $\zeta$ and, hence, to establish ''full-counting statistics'' of PC in our problem.
Real-time correlators might be obtained by analytic continuation.

For the sake of definiteness below we will set
$U(\theta)=U_0(1-\cos(\kappa\theta))$, i.e.
we will assume that the particle
on a 1D ring is moving in a periodic
potential with the distance $2\pi/\kappa$ between the adjacent
minima. The potential barriers
between these minima will be assumed high, $U_0\gg \kappa^2/(MR^2)$, in which case the
particle moves around the ring due to hopping from
one minimum to another.
Semiclassically, these hops are
described by multi-instanton trajectories
$\Theta(\tau)=\sum\limits_j\nu_j\tilde\theta(\tau-\tau_j)$ with
$\nu_j=\pm 1$,
which dominate the path integral (\ref{genf}). Here
$\tilde\theta(\tau)=4\arctan(e^{\Omega\tau})/\kappa$ is
well known kink solution, describing the particle tunneling with
the amplitude $\Delta /2=4(\Omega U_0/\pi )^{1/2}e^{-\frac{8U_0}{\Omega}}$,
where $\Omega =\kappa \sqrt{U_0/(MR^2)}$. Substituting the
trajectories $\Theta(\tau)$ into  Eq. (\ref{genf}) and performing Gaussian integration we get
 \begin{equation}
 \mathcal Z[\zeta ]=\mathcal Z_{0}[\zeta]\sum\limits_{k=1}^\kappa
 e^{\Delta\int\limits_0^\beta d\tau\cos\left(\frac{2\pi(\phi_x-k)}{\kappa}
 +\int\limits_0^\beta \zeta
 (\tau_1)\dot{\tilde\theta}(\tau-\tau_1)d\tau_1\right)}.
 \label{genfun}
 \end{equation}
Here $\mathcal Z_{0}[\zeta]$ is the generation functional for a harmonic oscillator. Employing the above results at $T\ll\Omega$ we obtain
\begin{equation}
 S_\omega=4\pi\mathcal P\delta(\omega)+\frac{e^2\Omega^3}{4\pi\kappa^2U_0\mathcal Z}
 \sum\limits_{k=1}^\kappa e^{\beta\Delta\cos\left(\frac{2\pi(\phi_x-k)}{\kappa}\right)}
\left(\delta(\omega-\Omega-\epsilon_k)+\delta(\omega+\Omega+\epsilon_k)\right),
\label{Sw}
\end{equation} where we defined $
\epsilon_k=(32U_0\Delta/\Omega)\cos\left(2\pi(\phi_x-k)/\kappa\right)$ and
 \begin{equation}
\mathcal P= \frac{e^2\Delta^2}{\kappa^2}\frac{\sum\limits_{k=1}^\kappa\sin^2\left(\frac{2\pi(\phi_x-k)}{\kappa}\right)
e^{\beta\Delta\cos\left(\frac{2\pi(\phi_x-k)}{\kappa}\right)}}{\sum\limits_{k=1}^\kappa
e^{\beta\Delta\cos\left(\frac{2\pi(\phi_x-k)}{\kappa}\right)}}
-\frac{e^2\Delta^2}{\kappa^2}\left(\frac{\sum\limits_{k=1}^\kappa\sin\left(\frac{2\pi(\phi_x-k)}{\kappa}\right)
e^{\beta\Delta\cos\left(\frac{2\pi(\phi_x-k)}{\kappa}\right)}}{\sum\limits_{k=1}^\kappa
e^{\beta\Delta\cos\left(\frac{2\pi(\phi_x-k)}{\kappa}\right)}}\right)^2
\label{pppp}
\end{equation}
As we already discussed, in the case of many channel rings or an ensemble of rings with varying parameter $U_0$ PC noise power consists of many close peaks, so that the noise spectrum gets smoother, as it is shown in Fig. 1a.
\begin{figure}
\begin{center}
a)\includegraphics[width=2.2in]{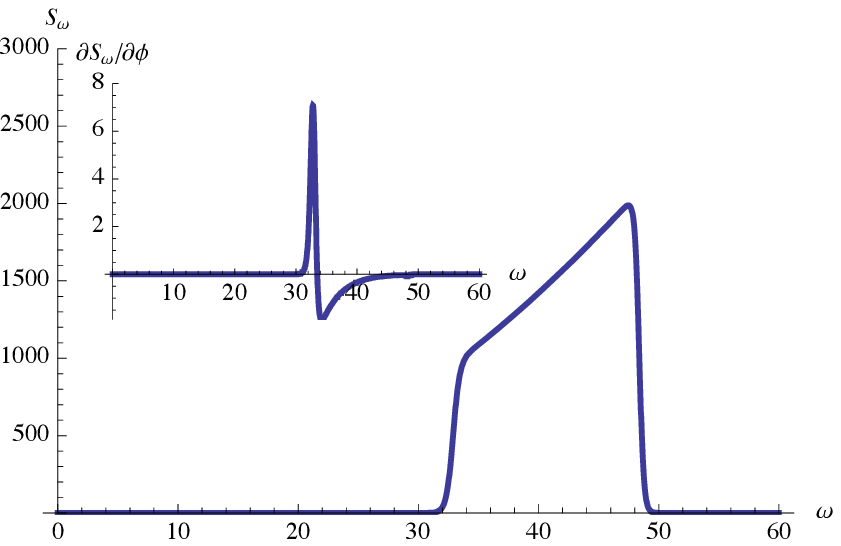}\qquad\qquad
b)\includegraphics[width=2.2in]{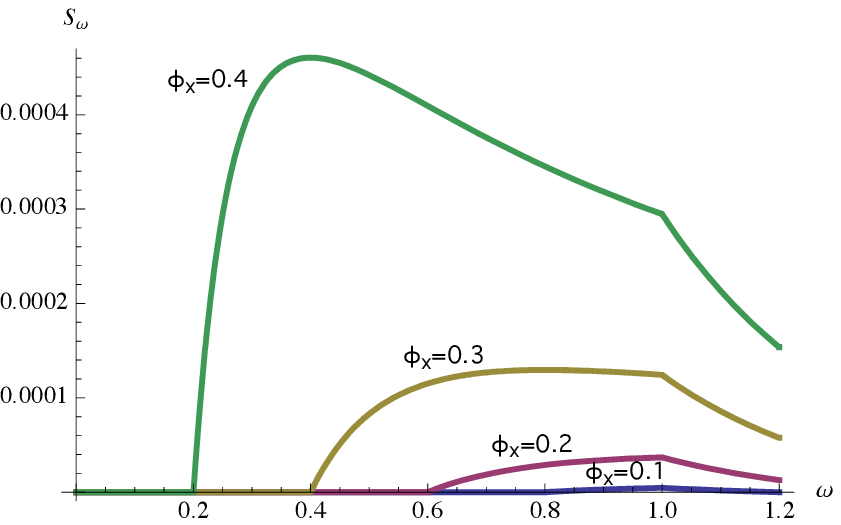}
\end{center}
\label{fig1}
\caption{(a)Zero temperature PC noise spectrum $S_\omega$
(arbitrary units) and its derivative with respect to the
    flux $\partial S_\omega/\partial \phi$ (arbitrary units) as
functions of $\omega$ (measured in units of $1/2MR^2$) for an ensemble of
rings (or for a ring containing many independent channels) with $U_0$
uniformly distributed within the interval from $30/MR^2$ to $65/MR^2$.(b) Zero temperature PC noise spectrum $S_\omega$ (arbitrary units)
was a function of $\omega$ (measured in units of $1/(2MR^2)$) for a dissipative system (\ref{diss}) for different fluxes $\phi_x$.}
\end{figure}

Our second example is a particle interacting with the dissipative bath formed by electrons in a disordered metal. In this case the action takes the form \cite{Paco,GHZ}
\begin{equation} S[\theta]=\frac{MR^2}{2}\int_0^\beta d\tau\left(\frac{\partial\theta}{\partial\tau}\right)^2+\alpha\int_0^\beta d\tau\int_0^\beta d\tau'\frac{\pi^2 T^2 K(\theta(\tau)-\theta(\tau'))}{\sin^2(\pi T(\tau-\tau'))}
\label{diss}
\end{equation}
Here $K(z)=1-\frac{1}{\sqrt{4r^2\sin^2(z/2)+1}}$, $\alpha=3/(8 k^2_F l^2)$ and $r=R/l$, $l$ is the electron mean free path. Below we will employ the perturbation in $\alpha$ technique developed by Golubev and one of the present authors \cite{GZ94}. Following this approach let us express the partition function in the following form:
\begin{equation}
 \mathcal Z=\sum\limits_{n=-\infty}^\infty Z[\beta,\phi_x-n]
\end{equation}
where the Laplace transform of $Z$ is defined as
\begin{equation}
 Z_p[q]=\int\limits_0^\infty d\tau e^{-p\tau} Z[\tau,q]=\frac{1}{p+q^2/(2MR^2)-\Sigma_p[q]}
\end{equation}
Here $\Sigma_p[q]$ is the self-energy defined as a sum of all irreducible diagrams.  Here  we proceed perturbatively  taking into account only the simplest one loop diagram. We also restrict our analysis to zero temperature limit. In that case the self-energy reads
 \begin{eqnarray}
\Sigma_p[q]=\frac{\alpha }{2}\sum\limits_{n=1}^r\alpha_n \sum\limits_{s=\pm1}(p+(q+sn)^2/(2MR^2))\ln(p+(q+sn)^2/(2MR^2))-\nonumber\\-\alpha\sum\limits_{n=1}^r\alpha_n(p+(q^2+n^2)/(2MR^2))\ln(p+(q^2+n^2)/(2MR^2))
\end{eqnarray}
where $\alpha_n=\frac{2}{\pi r}\ln(r/n)$. Making use of this expression it is straightforward to derive the current-current correlator. Neglecting vertex corrections we obtain
\begin{equation}
\langle\dot \theta(0)\dot\theta(\tau) \rangle=-\frac{1}{(MR^2)^2}\frac{\sum\limits_{n=-\infty}^\infty (n-\phi_x)^2 Z[\tau,n-\phi_x]Z[\beta-\tau,n-\phi_x]}{\sum\limits_{n=-\infty}^\infty Z[\beta,n-\phi_x]}
\end{equation}
After analytic continuation to real time we arrive at the following expression for zero-temperature PC noise power:
\begin{equation}
S_\omega=-\frac{e^2\phi_x^2}{2\pi^2 (MR^2)^2}\left(\rho_{\phi_x^2/(2MR^2)+\omega}[-\phi_x]+\rho_{\phi_x^2/(2MR^2)-\omega}[-\phi_x]\right)
\end{equation}
where $\rho_p[q]=\Im Z_{-p+i0}[q]$. This result is depicted in Fig. 1b.

In summary,  we investigated equilibrium fluctuations of
persistent current in nanorings and demonstrated that these
fluctuations do not vanish even at $T=0$ provided the current
operator does not commute with the total Hamiltonian of the
problem. A specific feature of PC noise is its
quantum coherent nature implying that the noise spectrum
can be tuned by an external magnetic flux inside the ring.
We believe that the key features captured by our
analysis will survive also in other models and can be verified in future
experiments.


\section*{References}
\begin{thebibliography}{9}
\bibitem{GZ98} Golubev D S and Zaikin A D 1998 {\it Physica B} {\bf 255} 164
\bibitem{Buttiker} Cedraschi P,  Ponomarenko V V and B\"uttiker M
2000 {\it Phys. Rev. Lett.} {\bf 84} 346
\bibitem{Paco} Guinea F 2002 {\it Phys. Rev. B} {\bf 65} 205317
\bibitem{GHZ} Golubev D S, Herrero C P and Zaikin A D 2003 {\it Europhys. Lett.} {\bf 63} 426
\bibitem{HlD} Horovitz B and Le Doussal P 2006 {\it Phys. Rev. B} {\bf 74} 073104
\bibitem{SZ} Semenov A G and Zaikin A D 2009 {\it Phys. Rev. B} {\bf 80} 155312
\bibitem{AGZ} Arutyunov K Yu,  Golubev D S and Zaikin A D 2008 {\it Phys. Rep.} {\bf 464} 1
\bibitem{GZ94} Golubev D S and Zaikin A D 1994 {\it Phys. Rev. B} {\bf 50}8736


\end{thebibliography}
\end{document}